\documentclass[sigconf,nonacm]{acmart}

\usepackage[utf8]{inputenc}
\usepackage[english]{babel}
\usepackage{hyperref}
\usepackage{amsmath}
\usepackage{amsfonts}
\usepackage{graphicx}
\usepackage{verbatim}
\usepackage{algorithm}
\usepackage{algpseudocode}
\usepackage{appendix}
\usepackage{xspace}
\usepackage[normalem]{ulem}
\usepackage{multicol}

\newcommand{\Put}{{\sf PUT}\xspace}
\newcommand{\Take}{{\sf TAKE}\xspace}
\newcommand{\compete}{{\sf compete}\xspace}
\newcommand{\RMW}{{\sf RMW}\xspace}
\newcommand{\CAS}{{\sf CAS}\xspace}
\newcommand{\FAI}{{\sf FAI}\xspace}
\newcommand{\R}{{\sf READ}\xspace}
\newcommand{\W}{{\sf WRITE}\xspace}
\newcommand{\LL}{{\sf LL}\xspace}
\newcommand{\IC}{{\sf IC}\xspace}
\newcommand{\SC}{{\sf SC}\xspace}
\newcommand{\SWAP}{{\sf SWAP}\xspace}
\newcommand{\Enq}{{\sf ENQ}\xspace}
\newcommand{\Deq}{{\sf DEQ}\xspace}
\newcommand{\ok}{{\sf OK}\xspace}
\newcommand{\closed}{{\sf CLOSED}\xspace}
\newcommand{\true}{{\sf true}\xspace}
\newcommand{\false}{{\sf false}\xspace}
\newcommand{\epty}{{\sf EMPTY}\xspace}
\newcommand{\full}{{\sf FULL}\xspace}
\newcommand{\open}{{\sf OPEN}\xspace}
\newcommand{\op}{{\sf op}\xspace}

\newcounter{linecounter}
\newcommand{\linenumbering}{\ifthenelse{\value{linecounter}<10}{(0\arabic{linecounter})}{(\arabic{linecounter})}}
\renewcommand{\line}[1]{\refstepcounter{linecounter}\label{#1}\linenumbering}
\newcommand{\resetline}[1]{\setcounter{linecounter}{0}#1}

\setcopyright{none}
\copyrightyear{2022}
\acmYear{2022}
\acmDOI{}
\acmConference[]{}{}{}
\acmBooktitle{}
\acmPrice{}
\acmISBN{}

\begin{document}

\title{Modular Baskets Queue}

\author{Armando Casta\~neda}
\email{armando.castaneda@im.unam.mx}
\affiliation{%
  \institution{Instituto de Matem\'aticas, UNAM}
  \city{Ciudad de M\'exico}         
  \country{M\'exico}   
}

\author{Miguel Pi\~na}
\email{miguel_pinia@ciencias.unam.mx }
\affiliation{%
  \institution{Faculad de Ciencias, UNAM}
  \city{Ciudad de M\'exico}         
  \country{M\'exico}   
}

\begin{abstract}
A modular version of the baskets queue of Hoffman,
Shalev and Shavit
is presented.
It manipulates the head and tail using a novel
object called \emph{load-link/increment-conditional},
which can be implemented using only \R/\W instructions, and
admits implementations that spread contention.
This suggests that there might be an alternative to the seemingly inherent
bottleneck in previous queue implementations that manipulate the head and the tail
using \emph{read-modify-write} instructions over a single shared register.
\end{abstract}


\maketitle


\section{Introduction}

Concurrent multi-producer/multi-consumer FIFO queues are fundamental shared data structures,
ubiquitous in all sorts of systems. For over more than three decades,
several concurrent queue shared-memory implementations
have been proposed. Despite these efforts,
even state-of-the-art concurrent queue algorithms scale poorly,
namely, as the number of cores grows, the latency of queue operations grow at least
linearly on the number of cores.

One of main the reasons of the poor scalability is the high contention in the \emph{read-modify-write} (\RMW)
instructions, such as \emph{compare-and-set} (\CAS) or \emph{fetch-and-increment} (\FAI),
that manipulate the head and the tail~\cite{FatourouK11,FatourouK12,HoffmanSS07,KoganP11,Ladan-MozesS08,
MichaelS96,MilmanKLLP18,MorrisonA13,OstrovskyM20,YangM16}. The latency of any contended such instruction
is linear in the number of contending cores, since every instruction acquires exclusive ownership
of its location’s cache line.
The best known queue implementations~\cite{MorrisonA13,YangM16} exploit the semantics
of the \FAI instruction, that \emph{do not fail} and hence \emph{always make progress}.
In many queue implementations, a queue operation \emph{retries} a failed \CAS until
it succeeds~\cite{FatourouK11,FatourouK12,KoganP11,Ladan-MozesS08,MichaelS96,MilmanKLLP18}.
An approach that lies in the middle is that of the \emph{baskets queue}~\cite{HoffmanSS07},
where a failed \CAS in an enqueue operation implies concurrency with other enqueue operations,
and hence the items of all these operations do not need to be ordered, instead they are
stored in a \emph{basket}, where the items can be dequeued in any order.
To overcome this seemingly inherent bottleneck, it has been recently proposed
a \CAS implementation from \emph{hardware transactional memory}, that exhibits better
performance that the usual \CAS~\cite{OstrovskyM20}.

In this ongoing project, we observe that \RMW instructions are not needed to consistently
manipulate the head or the tail. We believe that this observation might open the possibility
of concurrent queue implementations with better scalability.
Concretely, we present a \emph{modular} baskets queue algorithm,
based on a novel object that we call
\emph{load-link/increment-conditional} (\LL/\IC)
that suffices for manipulating the head and the tail of the queue.
\LL/\IC admits implementations that spread contention and use only simple \R/\W instructions.
\LL/\IC is a similar to \LL/\SC, with the difference
that \IC, if successful, only increments the current value of the linked register.
The modular baskets queue stands for its simplicity, with a simple correctness proof.



\section{The modular basket queue}
\label{sec-basket-queue}

\paragraph{Model of computation.}
We consider the standard shared memory model~\cite{HerlihyW90} with
$n \geq 2$ \emph{asynchronous} processes that communicate using \emph{atomic} instructions that modify the
contents of the shared memory; the instructions range from
simple \R and \W, to more complex \RMW instructions
such as \FAI and \CAS.
For simplicity, the baskets queue algorithm is presented using an infinite shared array~\footnote{An infinite array can be implemented using a linked list whose nodes contain arrays
of finite size; the list grows on
demand during an execution, each node appended to the list using \CAS
to maintain consistency.}.
We consider the \emph{wait-free}~\cite{Herlihy91} and
\emph{lock-free}~\cite{HerlihyS11}
progress conditions, and \emph{linearizability}~\cite{HerlihyW90} as
consistency condition.


\begin{algorithm}
  \caption{The modular baskets queue.} \label{basket-queue}
\begin{minipage}[t]{180mm} \footnotesize
\renewcommand{\baselinestretch}{2.5} \resetline
\begin{tabbing} aaaa\=aa\=aa\=aa\=aa\=aa\=aa\=\kill 

{\bf Shared Variables:}\\
\> \(A[0, 1, \ldots] =\) \textit{infinite array of basket objects}\\
\>  $HEAD, TAIL =$  \textit{\LL/\IC objects initialized to 0} \\ \\

{\bf Operation} $\Enq(x)$: \\
\line{A01} \> {\bf while \(\true\) do}\\
\line{A02} \> \> $tail = TAIL.\LL()$ \\
\line{A03} \> \> {\bf if $A[tail].\Put(x) == \ok$ then} \\
\line{A04} \> \> \> $TAIL.\IC()$ \\
\line{A05} \> \> \> \textbf{return} $\ok$\\
\line{A06} \> \> {\bf endif} \\
\line{A07} \> \> $TAIL.\IC()$ \\
\line{A08} \> {\bf endwhile}\\
{\bf end \Enq} \\ \\

{\bf Operation} $\Deq()$: \\
\line{A09} \> $head = HEAD.\LL()$ \\
\line{A10} \> $tail = TAIL.\LL()$ \\
\line{A11} \> {\bf while \(\true\) do}\\
\line{A12} \> \> {\bf if $head < tail$ then}\\
\line{A13} \> \> \> $x = A[head].\Take()$\\
\line{A14} \> \> \> {\bf if $x \neq \closed$ then return $x$ endif}\\
\line{A15} \> \> \> $HEAD.\IC()$ \\
\line{A16} \> \> {\bf endif} \\
\line{A17} \> \> $head' = HEAD.\LL()$ \\
\line{A18} \> \> $tail' = TAIL.\LL()$\\
\line{A19} \> \> {\bf if $head == head' == tail' == tail$ then return $\epty$ endif}\\
\line{A20} \> \> $head = head'$ \\
\line{A21} \> \> $tail = tail'$\\
\line{A22} \> {\bf endwhile}\\
{\bf end \Deq}
\end{tabbing}
\end{minipage}
\end{algorithm}


\paragraph{The algorithm.}
The modular baskets queue appears in Algorithm~\ref{basket-queue}.
It is based on two concurrent objects: baskets and \LL/\IC.
Roughly speaking, the baskets store groups of enqueued items that can be taken dequeued in any order,
while two \LL/\IC objects store the head and the tail of the queue.

The sequential specification of a \emph{basket of capacity $K$}, or \emph{$K$-basket},
satisfies the following properties, assuming the state of the object
is a pair $(S,C)$, initialized to $(\emptyset, 0)$:

(1) $\Put(x)$. Non-deterministically picks between returning \full (regardless of the state), and doing: If $C = K$, then return \full, else do $S = S \cup \{x\}$, $C = C +1$ and return $\ok$.

(2) $\Take()$. If $S \neq \emptyset$, then do $S = S \setminus \{x\}$ and return $x$, for some $x \in S$, else do $C = K$ and return \closed.

The baskets in the original baskets queue~\cite{HoffmanSS07}
were defined only \emph{implicitly}. Recently, baskets were explicitly defined in~\cite{OstrovskyM20}.
Our basket specification provides stronger guarantees,
being the main difference the following one.
In~\cite{OstrovskyM20}, there is a {\sf basket\_empty} operation that
can return either
\true or \false if the basket is not empty, i.e. it allows false negatives.
The \Take operation of our specification mixes the
functionality of
{\sf basket\_empty} and {\sf basket\_extract},
as if it returns \closed, no item will ever be put or
taken from the basket.

The specification of \LL/\IC satisfies the next properties,
where the state of the object is an integer $R$, initialized to 0, and assuming that any process invokes \IC only if it has invoked \LL before:
\begin{enumerate}
\item $\LL()$: Returns the current value in $R$.
\item $\IC()$: If $R$ has not been increment since the last \LL of the invoking process,
then do $R = R + 1$; in any case return \ok.
\end{enumerate}

\begin{theorem}
In Algorithm~\ref{basket-queue}, if the objects in $A$, and $HEAD$ and $TAIL$ objects are linearizable and wait-free,
then the algorithm is a linearizable lock-free implementation of a concurrent queue.
\end{theorem}

\begin{proof}
Since all shared objects are wait-free, every step of the implementation
completes. Note that every time a \Deq/\Enq operation completes
a while loop (hence without returning), an \Enq (resp. a \Deq) operation successfully
puts (resp. takes) an item in (resp. from) a basket.
Thus, in an infinite execution, if a \Deq/\Enq operation takes infinitely many steps,
infinitely many \Deq/\Enq operations terminate. Hence the implementation is lock-free.

To prove that the algorithm is linearizable, we consider the aspect-oriented linearizability
proof framework in~\cite{HenzingerSV13}. Assuming that every item is enqueued at most once,
it states that a queue implementation is lineairizable if each of its finite
executions is \emph{free} of four violation. We enumerate the violations and argue that
every execution of the algorithm is free of them.

VFresh: A \Deq operation returns an item not previously inserted by any \Enq operation.
Clearly, \Deq operations return
items that were previously put in the baskets, and \Enq operations put items in
the baskets. Thus, each execution is free of VFresh.

VRepeat: Two \Deq operations return the item inserted by the same \Enq operation.
The specification of the basket directly implies that every
execution is free of VRepeat.

VOrd: Two items are enqueued in a certain order, and a \Deq returns the later item before
any \Deq of the earlier item starts.
\LL/\IC guarantees that if an \Enq operation
enqueues an item, say $x$, and then a later \Enq operation enqueues
another item, say $y$, then $x$ and $y$ are inserted in baskets
$A[i]$ and $A[j]$, with $i < j$. Then, $x$ is dequeued first
because \Deq operations scan $A$ in index-ascending order.
Thus, every execution is free of VOrd.

VWit: A \Deq operation returning \epty even though the queue is never logically empty
during the execution of the \Deq operation.
An item is logically in the queue if it is in a basket $A[i]$ and $i < TAIL$.
When a \Deq operation returns \epty, there is a point in time where
no basket in $A[0, 1, \hdots, TAIL-1]$ contains an item, and hence
the queue is logically empty
(it might however be the case that $A[TAIL]$ does contain an item at that moment).
Hence every execution is free of VWit.
\end{proof}


The scalability of the algorithm depends on the scalability
of concrete implementations of \LL/\IC and basket
that it is instantiated with.
We propose wait-free implementations of each of the objects.

\paragraph{\LL/\IC implementations.} Let $p$ denote the process that invokes an operation.

\paragraph{A CAS-based implementation.}
It uses a shared register $R$ initialized to 0.
\LL first reads $R$ and stores the value
in a persistent variable $r_p$ of $p$, and then returns $r_p$.
\IC first reads $R$ and if that value is equal to $r_p$,
then it performs $\CAS(R, r_p, r_p+1)$; in any it case returns \ok.

\begin{theorem}
The CAS-based \LL/\IC implementation just described
is linearizable and wait-free.
\end{theorem}

\begin{proof}[Proof sketch.]
The algorithm is obviously wait-free.
For the linearizability proof, consider any finite execution $E$ with no pending operations.
We define the following linearization points.
The linearization point of an \LL operation is when it reads $R$.
If an \IC operation performs a \CAS, it is
linearized at that step, otherwise it is linearized when it reads $R$.
Let $S_t$ be the sequential execution induced by the first $t$ linearization
points of $E$, reading its steps in index-ascending order.
By induction on $t$, it can be shown that $S_T$ is a sequential execution of \LL/\IC,
where $T$ is the number of operations in $E$.
The main observation is that that if there a successful \CAS before
the \CAS of an \IC operation of process $p$, then the contents of $R$
is different from the value $p$ reads in its previous \LL operation.
\end{proof}

\paragraph{A \R/\W implementation.}
It uses a shared array $M$ with $n$ entries initialized to 0.
\LL first reads all entries of $M$ (in some order) and stores the maximum value in a persistent variable $max_p$ of $p$,
and then returns $max_p$.
\IC first reads all entries of $M$, and if the maximum among those values is equal to $max_p$,
it performs $\W(M[p], max_p + 1)$; in any it case returns \ok.

\begin{theorem}
The \R/\W-based \LL/\IC implementation just described 
is linearizable and wait-free.
\end{theorem}

\begin{proof}[Proof sketch]
The algorithm is obviously wait-free. We next argue that each of its executions is linearizable.

Consider any finite execution of the algorithm with no pending operations.
To make or argument simple,
let us suppose that there is a \emph{fictitious} \IC operation that atomically writes 0
in all entries of $M$ at the very beginning of the execution.

Each \IC operation is linearized at its last
step. Thus, an \IC that writes, is linearized at its \W step, and an \IC that does not
write is linearized at its last \R step.
Let $MAX$ be the maximum value in the shared array $M$ at the end of the execution.
For every $R \in \{0,1, \hdots, MAX\}$, let $\IC_R$ be the \IC operation that writes $R$ for
the first time in $M$.

We will linearize every \LL operation that returns value $R \in \{0, 1, \hdots, MAX-1\}$
at one of its step, and argue that this step is between $\IC_R$ and $\IC_{R+1}$.
This will induce a sequential execution that respect the real-time order
and is a sequential execution of \LL/\IC, and hence a linearization.

Let \op denote any \LL that returns $R \in \{0, 1, \hdots, MAX-1\}$
and let $e$ denote its \R step that reads $R$ for the first time.
Observe that $IC_R$ has been linearized when $e$ happens in the execution.
We have two cases:
\begin{enumerate}
    \item If the shared memory $M$ does not contain a value $> R$ when $e$ occurs
    (hence no $IC_{R'}$ with $R' > R$ has been linearized when $e$ occurs),
    then \op is linearized at $e$.

    \item If the shared memory $M$ does contain a value $> R$ when $e$ occurs,
    then \op is linearized as follows.
    Let $M[j]$ be the entry that is read at step $e$. Note that this case can happen
    if and only if some entries in the range $M[0, \hdots, j-1]$
    contain values $> R$ when $e$ happens
    (and hence some $IC_{R'}$ with $R' > R$ have been linearized when $e$ occurs).
    Moreover, it can be shown that the value $R+1$ is written in an entry in the range
    $M[0, \hdots, j-1]$ at some time between the invocation of \op and $e$.
    Let $i \in \{0, \hdots, j-1\}$ be the index of the entry where it is written
    $R+1$ for the first time. Then, \op is linearized right before $R+1$ is
    written in $M[i]$ (and hence before $\IC_{R+1}$).
\end{enumerate}

\end{proof}

\paragraph{A mixed implementation.}
It uses a shared array $M$ with $K < n$ entries initialized to 0.
\LL reads all entries of $M$ and stores the maximum value and its index
in persistent variables $max_p$ and $indmax_p$ of $p$,
and returns $max_p$.
\IC non-deterministically picks an index
$pos \in \{0,1,\hdots,K-1\} \setminus \{indmax_p\}$.
If $M[pos]$ contains a value $x$ less than $max_p+1$, then it performs
$\CAS(M[pos], x, max_p+1)$; if the \CAS is successful, it returns \ok.
Otherwise, it reads the value in $M[indmax_p]$, and if it is equal to $max_p$,
then it performs $\CAS(M[indmax_p], max_p, max_p+1)$;
in any it case returns~\ok.

\begin{theorem}
The mixed implementation just described
is linearizable and wait-free.
\end{theorem}

\begin{proof}[Proof sketch.]
The algorithm is obviously wait-free.
The linearizability proof is nearly the same as the one in the previous theorem proof;
the only difference is that each \IC operation is linearized at its last
step, either a \CAS (successful or not) or a \R.
\end{proof}

\begin{algorithm}
  \caption{$K$-basket from \FAI and \SWAP.}\label{basket-1}

  \begin{minipage}[t]{180mm} \footnotesize
\renewcommand{\baselinestretch}{2.5} \resetline
\begin{tabbing} aaaa\=aa\=aa\=aa\=aa\=aa\=aa\=\kill 

{\bf Shared Variables:}\\
\> \(A[0, 1, \ldots, K-1] = [\bot, \bot, \ldots, \bot]\) \\
\> \(PUTS, TAKES = 0\) \\
\> \(STATE = \open\) \\ \\

{\bf Operation} $\Put(x)$: \\
\line{B01} \> {\bf while \(\true\) do}\\
\line{B02} \> \> \(state = \R(STATE)\) \\
\line{B03} \> \> \(puts = \R(PUTS)\) \\
\line{B04} \> \> {\bf if \(state == \closed\) or \(puts \geq K\) then return \full} \\
\line{B05} \> \> {\bf else} \\
\line{B06} \> \> \> \(puts = \FAI(PUTS)\) \\
\line{B07} \> \> \> {\bf if \(puts \geq K\) then return \full} \\
\line{B08} \> \> \> {\bf else if \(\SWAP(A[puts], x) == \bot\) then return \ok endif} \\
\line{B09} \> \> {\bf endif} \\
\line{B10} \> {\bf endwhile}\\
{\bf end \Put} \\ \\

{\bf Operation} $\Take()$: \\
\line{B11} \> {\bf while \(\true\) do}\\
\line{B12} \> \> \(state = \R(STATE)\) \\
\line{B13} \> \> \(takes = \R(TAKES)\) \\
\line{B14} \> \> {\bf if \(state == \closed\) or \(takes \geq K\) then return \closed} \\
\line{B15} \> \> {\bf else} \\
\line{B16} \> \> \> \(takes = \FAI(TAKES)\) \\
\line{B17} \> \> \> {\bf if \(takes \geq K\) then} \\
\line{B18} \> \> \> \> \(\W(STATE, \closed)\) \\
\line{B19} \> \> \> \> \textbf{return} \closed \\
\line{B20} \> \> \> {\bf else} \\
\line{B21} \> \> \> \> \(x = \SWAP(A[puts], \top)\) \\
\line{B22} \> \> \> \> {\bf if \(x \neq \bot\) then return \(x\) endif} \\
\line{B23} \> \> \> {\bf endif} \\
\line{B24} \> \> {\bf endif} \\
\line{B25} \> {\bf endwhile}\\
{\bf end \Take}
\end{tabbing}
\end{minipage}
\end{algorithm}

\begin{algorithm}
  \caption{$n$-basket from \CAS. Let $p$ denote the invoking process.}\label{basket-2}
  \begin{minipage}[t]{180mm} \footnotesize
\renewcommand{\baselinestretch}{2.5} \resetline
\begin{tabbing} aaaa\=aa\=aa\=aa\=aa\=aa\=aa\=\kill 

{\bf Shared Variables:}\\
\> \(A[0, 1, \ldots, n-1] = [\bot, \bot, \ldots, \bot]\) \\
\> \(STATE = \open\) \\
{\bf Persistent Local Variables of $p$:}\\
\> \(takes_p = \{0, 1, \hdots, n-1\}\)\\ \\

{\bf Operation} $\Put(x)$: \\
\line{C01} \> {\bf if \(\R(STATE) == \closed\) then return \full} \\
\line{C02} \> {\bf else if \(\R(A[p]) == \bot \) then  } \\
\line{C03} \> \> {\bf if \(\CAS(A[p], \bot, x) \) then return \ok endif} \\
\line{C04} \> {\bf endif} \\
\line{C05} \> {\bf return \full}\\
{\bf end \Put} \\ \\

{\bf Function} $\compete(pos)$: \\
\line{C06} \> \(x = \R(A[pos])\) \\
\line{C07} \> {\bf if \(x == \top \) then return \(\top\)} \\
\line{C08} \> {\bf else if \( \CAS(A[pos], x, \top)  \) then return \(x\)  } \\
\line{C09} \> {\bf else return \(\bot\) endif} \\
{\bf end \compete} \\ \\

{\bf Operation} $\Take()$: \\
\line{C10} \> {\bf while \(\true\) do}\\
\line{C11} \> \> {\bf if \(\R(STATE) == \closed\) then return \closed} \\
\line{C12} \> \> {\bf else} \\
\line{C13} \> \> \> {\bf if \(p \in takes_p\) then \(pos = p\)} \\
\line{C14} \> \> \> {\bf else} \(pos = \hbox{any element of } takes_p\) {\bf endif} \\
\line{C15} \> \> \> \(takes_p =  takes_p \setminus \{pos\}\) \\
\line{C16} \> \> \> {\bf if \(takes_p == \emptyset\) then \(\W(STATE, \closed)\) endif} \\
\line{C17} \> \> \> \( x = \compete(pos) \) \\
\line{C13} \> \> \> {\bf if \(x \neq \bot, \top \) then return \(x\)} \\
\line{C14} \> \> \> {\bf else if \(x == \bot\) then} \\
\line{C19} \> \> \> \> \( x = \compete(pos) \) \\
\line{C22} \> \> \> \> {\bf if \(x \neq \bot, \top \) then return \(x\) endif} \\
\line{C24} \> \> \>{\bf endif} \\
\line{C24} \> \> {\bf endif} \\
\line{C25} \> {\bf endwhile}\\
{\bf end \Take}
\end{tabbing}
\end{minipage}
\end{algorithm}


\paragraph{Basket implementations.}
The basket implementations appear in Algorithms~\ref{basket-1} and~\ref{basket-2}.
5structure implementations of~\cite{HaasHHKLPSSV16}.

In the first implementation, the processes use \FAI to guarantee
that at most two "opposite" operations "compete" for the same location in the
shared array, which can be resolved with a \SWAP; the idea of this algorithm is
similar to the approach in the LCRQ algorithm~\cite{MorrisonA13}.

In the second implementation, each process has a dedicated location in the shared array where
it tries to put its item when it invokes \Put. When a process invokes \Take,
it first tries to take an item from its dedicated location, and
if it does not succeed, it randomly picks non-previously-picked location and
does the same, and repeats  until it takes an item or all locations have been cancelled.
Since several operations might "compete" for the same location, \CAS is needed.
This implementation is reminiscent to \emph{locally linearizable} generic data
structure implementations of~\cite{HaasHHKLPSSV16}.

\begin{theorem}
Algorithm~\ref{basket-1} is a wait-free linearizable implementation of a $K$-basket.
\end{theorem}

\begin{proof}[Proof sketch]
It is not hard to see that the algorithm is wait-free.

For the linearizability proof, given an entry $A[i]$, we will say that a \Put operation
\emph{successfully puts} its item in $A[i]$ if it gets $\bot$ when it performs \SWAP
on $A[i]$, and that a \Take operation \emph{successfully cancels} $A[i]$ if it gets
$\bot$ when it performs \SWAP on $A[i]$, otherwise (i.e. it gets a value distinct from $\bot$),
we say that the \Take operation \emph{successfully takes} an item from $A[i]$.

From the specification of \FAI, for every $A[i]$, at most one \Put operations tries to
successfully put its item in $A[i]$, and at most one \Take operation tries to either
successfully cancel $A[i]$ or successfully take an item from $A[i]$.
By the specification of \SWAP, if $A[i]$ is cancelled, no \Put operation successfully
puts an item in it and no \Take operation successfully takes an item from it.

Given any execution of the algorithm, the operations are linearized as follows.
A \Put operation that successfully puts its item is linearizaed at its last \FAI
instruction before returning.
A \Take operation that successfully takes an item from $A[i]$,
is linearizaed right after the \Put operation that successfully
put its item in $A[i]$.
A \Put that returns \full is linearized at its return step,
and, similarly, a \Take that returns \closed is linearized at its return step.
Note that, in both cases, at that moment of the execution,
every entry of $A$ has been or will be either cancelled or a \Take operation
has or will successfully take an item from it.
It can be shown that these linearization points induce a valid linearization of the execution.
\end{proof}

\begin{theorem}
Algorithm~\ref{basket-2} is a wait-free linearizable implementation of an $n$-basket.
\end{theorem}

\begin{proof}[Proof sketch]
Clearly, \Put is wait-free. It is not difficult to see that \Take is wait-free too.

For the linearizability proof, given an entry $A[i]$, we will say that a \Put operation of
process $p$, \emph{successfully puts} its item in $A[p]$ if its \CAS is successful.
A \Take operation \emph{successfully cancels} $A[i]$ if its $\CAS(A[i], x, \top)$
(in the \compete function) is successful, with $x$ being $\bot$;
and it \emph{successfully takes} an item from $A[i]$
if its $\CAS(A[i], x, \top)$
(in the \compete function) is successful, with $x$ being distinct to $\bot$ and $\top$.

The linearizability proof is similar to the linearizabiloty proof in the previous
theorem, with the following main differences.
(1) If a \Put operation returns \full, it can be the case that some of the other
entries of $A$ will never be cancelled or store an item; the response
of the \Put operation is however correct because the sequential specification
of $n$-basket allows \Put to return \full in any state of the object.
(2) Several \Take operations might try to either successfully cancel the same
entry $A[i]$ or successfully take an item from it; this is not a problem
because the specification of \CAS guarantees that at most one succeeds in
doing this.

Given any execution of the algorithm, the operations are linearized as follows.
A \Put operation that successfully puts its item is linearizaed at its (successful) \CAS.
A \Take operation that successfully takes an item from $A[i]$,
is linearizaed right after the \Put operation that successfully
put its item in $A[i]$.
A \Put that returns \full is linearized at its return step,
and, similarly, a \Take that returns \closed is linearized at its return step.
Note that at the moment of the execution a \Take that returns \closed,
every entry of $A$ has been either cancelled or a \Take operation
has successfully take an item from it.
It can be shown that these linearization points induce a valid linearization of the execution.
\end{proof}

\section{Preliminary Experiment}


The three proposed \LL/\IC implementations were evaluated, and an
implementation where the processes perform \FAI over the same
register. The latter implementation was considered as the best
concurrent queues manipulate the head using \FAI.  The experiment was
performed in an AMD Threadripper 3970X machine with 32 cores, each
multiplexing 2 hardware threads, allowing 64 threads in total; each
core has private L1 and L2 caches, and shares an L3 cache.


In the \LL/\IC implementations, each thread calls a \LL followed by
\IC, and, between each call to these methods, work of some length is
executed to avoid artificial long run scenarios (see for
example~\cite{YangM16}).  This work is a cycle with random increments,
one to five, where the limit of the cycle is a small number,
concretely \(25\) in the experiment. It was measured the time it took
each process to complete \(5\cdot10^{6}\) interspersed \LL and \IC,
with a respective random work; similarly, in the \FAI implementation,
each thread performed \(5\cdot10^{6}\) {\FAI}s with random work. The
\emph{false sharing} problem~\cite{BoloskyMichael93} was taken into
account in the array based \LL/\IC implementations (i.e. \R/\W
and the mixed one). The implementations with padding, for avoiding false sharing,
were not better than than implementations without padding, which are
the ones reported below.


\begin{figure}[ht]
  \centering
  \includegraphics[scale=0.17]{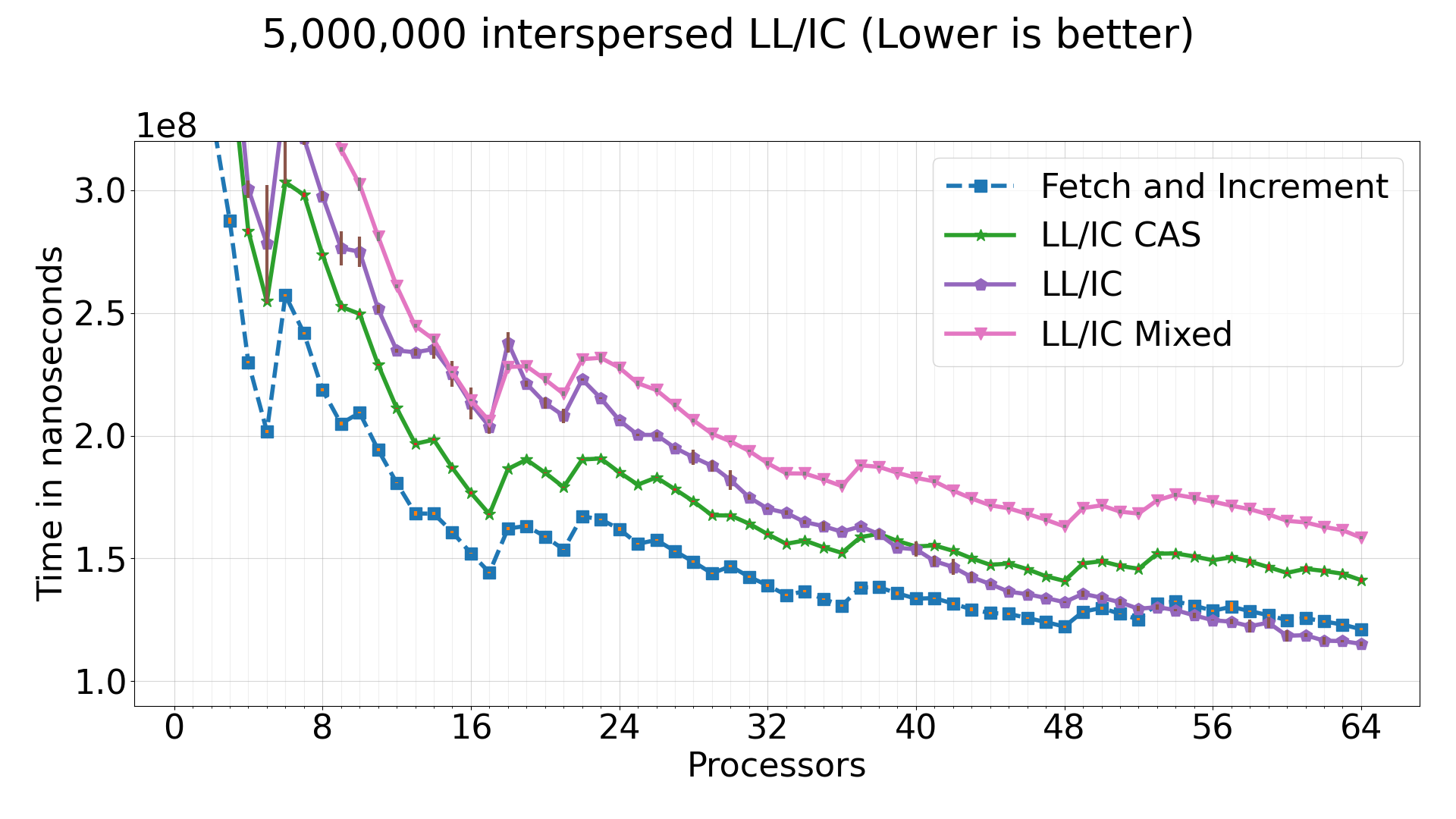}
  \caption{\label{fig:evaluation} Time to perform
    5,000,000 \LL/\IC interspersed operations per process.}
\end{figure}


Figure~\ref{fig:evaluation} shows the result of the experiment.  It
report averages of 5 executions from one to 64 threads, and error bars
indicating standard deviation.  The \FAI implementation had the best
performance, followed by the \CAS implementation of \LL/\IC.  The
\R/\W implementation of \LL/\IC improves its performance respect to
the previous two implementations, approaching and even being better
than the previous two implementations as the number of threads
increases.  An explanation is that contention is spread over the
entries of the array, reducing the number of hits to the same cache
line.  Finally, the mixed version of \LL/\IC, with $K = 2$, had the
worst performance but it is close to the \LL/\IC CAS implementation.

\section{Final remarks}

The next step of this ongoing project is finding implementations of
basket and \LL/\IC with good scalability, and compare the performance
of the resulting baskets queue with the known queue implementations.
It also might be worth to explore implementations of \LL/\IC using
hardware transactional memory. That approach was useful
in~\cite{OstrovskyM20} for boosting the scalability of \CAS
operations.

\bibliographystyle{ACM-Reference-Format}
\bibliography{references}


\begin{thebibliography}{16}


\ifx \showCODEN    \undefined \def \showCODEN     #1{\unskip}     \fi
\ifx \showDOI      \undefined \def \showDOI       #1{#1}\fi
\ifx \showISBNx    \undefined \def \showISBNx     #1{\unskip}     \fi
\ifx \showISBNxiii \undefined \def \showISBNxiii  #1{\unskip}     \fi
\ifx \showISSN     \undefined \def \showISSN      #1{\unskip}     \fi
\ifx \showLCCN     \undefined \def \showLCCN      #1{\unskip}     \fi
\ifx \shownote     \undefined \def \shownote      #1{#1}          \fi
\ifx \showarticletitle \undefined \def \showarticletitle #1{#1}   \fi
\ifx \showURL      \undefined \def \showURL       {\relax}        \fi
\providecommand\bibfield[2]{#2}
\providecommand\bibinfo[2]{#2}
\providecommand\natexlab[1]{#1}
\providecommand\showeprint[2][]{arXiv:#2}

\bibitem[\protect\citeauthoryear{Fatourou and Kallimanis}{Fatourou and
  Kallimanis}{2011}]%
        {FatourouK11}
\bibfield{author}{\bibinfo{person}{P. Fatourou} {and} \bibinfo{person}{N.~D.
  Kallimanis}.} \bibinfo{year}{2011}\natexlab{}.
\newblock \showarticletitle{A highly-efficient wait-free universal
  construction}. In \bibinfo{booktitle}{\emph{{SPAA} 2011)}}.
  \bibinfo{publisher}{{ACM}}, \bibinfo{pages}{325--334}.
\newblock


\bibitem[\protect\citeauthoryear{Fatourou and Kallimanis}{Fatourou and
  Kallimanis}{2012}]%
        {FatourouK12}
\bibfield{author}{\bibinfo{person}{P. Fatourou} {and} \bibinfo{person}{N.~D.
  Kallimanis}.} \bibinfo{year}{2012}\natexlab{}.
\newblock \showarticletitle{Revisiting the combining synchronization
  technique}. In \bibinfo{booktitle}{\emph{{PPOPP} 2012}}.
  \bibinfo{publisher}{{ACM}}, \bibinfo{pages}{257--266}.
\newblock


\bibitem[\protect\citeauthoryear{Haas, Henzinger, Holzer, Kirsch, Lippautz,
  Payer, Sezgin, Sokolova, and Veith}{Haas et~al\mbox{.}}{[n.d.]}]%
        {HaasHHKLPSSV16}
\bibfield{author}{\bibinfo{person}{A Haas}, \bibinfo{person}{T.~A. Henzinger},
  \bibinfo{person}{A. Holzer}, \bibinfo{person}{C.~M. Kirsch},
  \bibinfo{person}{M. Lippautz}, \bibinfo{person}{H. Payer},
  \bibinfo{person}{A. Sezgin}, \bibinfo{person}{A. Sokolova}, {and}
  \bibinfo{person}{H. Veith}.} \bibinfo{year}{[n.d.]}\natexlab{}.
\newblock \showarticletitle{Local Linearizability for Concurrent Container-Type
  Data Structures}. In \bibinfo{booktitle}{\emph{{CONCUR} 2016}}
  \emph{(\bibinfo{series}{LIPIcs}, Vol.~\bibinfo{volume}{59})}.
  \bibinfo{pages}{6:1--6:15}.
\newblock


\bibitem[\protect\citeauthoryear{Henzinger, Sezgin, and Vafeiadis}{Henzinger
  et~al\mbox{.}}{2013}]%
        {HenzingerSV13}
\bibfield{author}{\bibinfo{person}{T.~A. Henzinger}, \bibinfo{person}{A.
  Sezgin}, {and} \bibinfo{person}{V. Vafeiadis}.}
  \bibinfo{year}{2013}\natexlab{}.
\newblock \showarticletitle{Aspect-Oriented Linearizability Proofs}. In
  \bibinfo{booktitle}{\emph{{CONCUR} 2013}} \emph{(\bibinfo{series}{LNCS},
  Vol.~\bibinfo{volume}{8052})}. \bibinfo{publisher}{Springer},
  \bibinfo{pages}{242--256}.
\newblock


\bibitem[\protect\citeauthoryear{Herlihy}{Herlihy}{1991}]%
        {Herlihy91}
\bibfield{author}{\bibinfo{person}{M. Herlihy}.}
  \bibinfo{year}{1991}\natexlab{}.
\newblock \showarticletitle{Wait-Free Synchronization}.
\newblock \bibinfo{journal}{\emph{{ACM} Trans. Program. Lang. Syst.}}
  \bibinfo{volume}{13}, \bibinfo{number}{1} (\bibinfo{year}{1991}),
  \bibinfo{pages}{124--149}.
\newblock


\bibitem[\protect\citeauthoryear{Herlihy and Shavit}{Herlihy and
  Shavit}{2011}]%
        {HerlihyS11}
\bibfield{author}{\bibinfo{person}{M. Herlihy} {and} \bibinfo{person}{N.
  Shavit}.} \bibinfo{year}{2011}\natexlab{}.
\newblock \showarticletitle{On the Nature of Progress}. In
  \bibinfo{booktitle}{\emph{{OPODIS} 2011}} \emph{(\bibinfo{series}{LNCS},
  Vol.~\bibinfo{volume}{7109})}. \bibinfo{publisher}{Springer},
  \bibinfo{pages}{313--328}.
\newblock


\bibitem[\protect\citeauthoryear{Herlihy and Wing}{Herlihy and Wing}{1990}]%
        {HerlihyW90}
\bibfield{author}{\bibinfo{person}{M. Herlihy} {and} \bibinfo{person}{J.~M.
  Wing}.} \bibinfo{year}{1990}\natexlab{}.
\newblock \showarticletitle{Linearizability: {A} Correctness Condition for
  Concurrent Objects}.
\newblock \bibinfo{journal}{\emph{{ACM} Trans. Program. Lang. Syst.}}
  \bibinfo{volume}{12}, \bibinfo{number}{3} (\bibinfo{year}{1990}),
  \bibinfo{pages}{463--492}.
\newblock


\bibitem[\protect\citeauthoryear{Hoffman, Shalev, and Shavit}{Hoffman
  et~al\mbox{.}}{2007}]%
        {HoffmanSS07}
\bibfield{author}{\bibinfo{person}{M. Hoffman}, \bibinfo{person}{O. Shalev},
  {and} \bibinfo{person}{N. Shavit}.} \bibinfo{year}{2007}\natexlab{}.
\newblock \showarticletitle{The Baskets Queue}. In
  \bibinfo{booktitle}{\emph{{OPODIS} 2007}} \emph{(\bibinfo{series}{LNCS},
  Vol.~\bibinfo{volume}{4878})}. \bibinfo{publisher}{Springer},
  \bibinfo{pages}{401--414}.
\newblock


\bibitem[\protect\citeauthoryear{J. and Scott}{J. and Scott}{1993}]%
        {BoloskyMichael93}
\bibfield{author}{\bibinfo{person}{W.~Bolosky J.} {and} \bibinfo{person}{M.~L.
  Scott}.} \bibinfo{year}{1993}\natexlab{}.
\newblock \showarticletitle{False Sharing and Its Effect on Shared Memory
  Performance}. In \bibinfo{booktitle}{\emph{USENIX SEDMS 1993}}.
  \bibinfo{publisher}{USENIX Association}, \bibinfo{address}{USA},
  \bibinfo{pages}{3}.
\newblock


\bibitem[\protect\citeauthoryear{Kogan and Petrank}{Kogan and Petrank}{2011}]%
        {KoganP11}
\bibfield{author}{\bibinfo{person}{A. Kogan} {and} \bibinfo{person}{E.
  Petrank}.} \bibinfo{year}{2011}\natexlab{}.
\newblock \showarticletitle{Wait-free queues with multiple enqueuers and
  dequeuers}. In \bibinfo{booktitle}{\emph{{PPOPP} 2011}}.
  \bibinfo{publisher}{{ACM}}, \bibinfo{pages}{223--234}.
\newblock


\bibitem[\protect\citeauthoryear{Ladan{-}Mozes and Shavit}{Ladan{-}Mozes and
  Shavit}{2008}]%
        {Ladan-MozesS08}
\bibfield{author}{\bibinfo{person}{E. Ladan{-}Mozes} {and} \bibinfo{person}{N.
  Shavit}.} \bibinfo{year}{2008}\natexlab{}.
\newblock \showarticletitle{An optimistic approach to lock-free {FIFO} queues}.
\newblock \bibinfo{journal}{\emph{Distributed Computing}} \bibinfo{volume}{20},
  \bibinfo{number}{5} (\bibinfo{year}{2008}), \bibinfo{pages}{323--341}.
\newblock


\bibitem[\protect\citeauthoryear{Michael and Scott}{Michael and Scott}{1996}]%
        {MichaelS96}
\bibfield{author}{\bibinfo{person}{M.~M. Michael} {and} \bibinfo{person}{M.~L.
  Scott}.} \bibinfo{year}{1996}\natexlab{}.
\newblock \showarticletitle{Simple, Fast, and Practical Non-Blocking and
  Blocking Concurrent Queue Algorithms}. In \bibinfo{booktitle}{\emph{{PODC}
  1996}}. \bibinfo{publisher}{{ACM}}, \bibinfo{pages}{267--275}.
\newblock


\bibitem[\protect\citeauthoryear{Milman, Kogan, Lev, Luchangco, and
  Petrank}{Milman et~al\mbox{.}}{2018}]%
        {MilmanKLLP18}
\bibfield{author}{\bibinfo{person}{G. Milman}, \bibinfo{person}{A. Kogan},
  \bibinfo{person}{Y. Lev}, \bibinfo{person}{V. Luchangco}, {and}
  \bibinfo{person}{E. Petrank}.} \bibinfo{year}{2018}\natexlab{}.
\newblock \showarticletitle{{BQ:} {A} Lock-Free Queue with Batching}. In
  \bibinfo{booktitle}{\emph{{SPAA} 2018}}. \bibinfo{publisher}{{ACM}},
  \bibinfo{pages}{99--109}.
\newblock


\bibitem[\protect\citeauthoryear{Morrison and Afek}{Morrison and Afek}{2013}]%
        {MorrisonA13}
\bibfield{author}{\bibinfo{person}{A. Morrison} {and} \bibinfo{person}{Y.
  Afek}.} \bibinfo{year}{2013}\natexlab{}.
\newblock \showarticletitle{Fast concurrent queues for x86 processors}. In
  \bibinfo{booktitle}{\emph{{PPoPP} 2013}}. \bibinfo{publisher}{{ACM}},
  \bibinfo{pages}{103--112}.
\newblock


\bibitem[\protect\citeauthoryear{Ostrovsky and Morrison}{Ostrovsky and
  Morrison}{2020}]%
        {OstrovskyM20}
\bibfield{author}{\bibinfo{person}{O. Ostrovsky} {and} \bibinfo{person}{A.
  Morrison}.} \bibinfo{year}{2020}\natexlab{}.
\newblock \showarticletitle{Scaling concurrent queues by using {HTM} to profit
  from failed atomic operations}. In \bibinfo{booktitle}{\emph{{PPoPP} 2020}}.
  \bibinfo{publisher}{{ACM}}, \bibinfo{pages}{89--101}.
\newblock


\bibitem[\protect\citeauthoryear{Yang and Mellor{-}Crummey}{Yang and
  Mellor{-}Crummey}{2016}]%
        {YangM16}
\bibfield{author}{\bibinfo{person}{C. Yang} {and} \bibinfo{person}{J.~M.
  Mellor{-}Crummey}.} \bibinfo{year}{2016}\natexlab{}.
\newblock \showarticletitle{A wait-free queue as fast as fetch-and-add}. In
  \bibinfo{booktitle}{\emph{{PPoPP} 2016}}. \bibinfo{publisher}{{ACM}},
  \bibinfo{pages}{16:1--16:13}.
\newblock


\end{thebibliography}

\end{document}